\documentstyle[seceq]{ptptex}

\title{
Gauge Symmetry Reduction from the Extra Space ${\rm S}^1/{\rm Z}_2$
}

\author{
Yoshiharu {\sc Kawamura}\footnote{E-mail address:
haru@azusa.shinshu-u.ac.jp}
}

\inst{
Department of Physics, Shinshu University, Matsumoto 390-8621, Japan
}

\recdate{
April 5, 1999
}

\abst{
We study a mechanism of symmetry transition upon compactification of
a 5-dimensional field theory on $S^1/Z_2$.
The transition occurs unless all components in a multiplet 
of a symmetry group have a common $Z_2$ parity on $S^1/Z_2$.
This mechanism is applied to a reduction of $SU(5)$ gauge symmetry 
in grand unified theory, and phenomenological implications are discussed.
}

\begin{document}

\maketitle

\section{Introduction}

The study of physics beyond 4-dimensional (4D) space-time traces
back to the work by Kaluza and Klein.\cite{KK}
The exsitence of an extra space is essential in
superstring theory \cite{CY,Orb} and $M$-theory.\cite{HW}
Recently, there is a lot of interesting research of the phenomenological
and theoretical implications of an extra space with large radii
based on several motivations.\cite{largeR}

As a feature in theories with an extra compact space $K$,
symmetries in the system can change after compactification.
Here we give typical examples, where $K$ is 
an $n$-dimensional torus $T^n$.
Supersymmetry (SUSY) is, in general, enhanced
by a simple dimensional reduction.\cite{DR}
Also, SUSY can be broken by the choice of different
boundary conditions for bosons and fermions in supermultiplets.
This mechanism is called the Scherk-Schwarz mechanism.\cite{SS}
The other example is a gauge symmetry breaking
by the Hosotani mechanism.\cite{H,W}
Gauge symmetry is broken by the appearance of
non-integrable phase factors which are dynamical degrees of freedom
and are related to Wilson loops of gauge fields along 
compactified directions.

The phenomenon of symmetry transition leads to the idea that
symmetries in 4D low-energy theory are derived from
a high-energy theory with extra dimensions.
It is meaningful to study the relation between symmetries
and a compact space more carefully,
because the origin of symmetries in the Standard Model 
is not yet known.
As a compact space, the one-dimensional orbifold $S^1/Z_2$
has attracted our attention since the study of
heterotic M-theory.\cite{HW}
Starting from 5-dimensional (5D) SUSY model, 4D theory with $N=1$ SUSY
is derived through compactification on $S^1/Z_2$.\cite{MP,PQ}
Bulk fields have a $Z_2$ parity in a compact direction.
The reduction of SUSY originates from a non-universal $Z_2$ parity
assignment among component fields in supermultiplets.

In this paper, we focus on 5D field theories
and study a symmetry transition upon compactification on $S^1/Z_2$.
The above-mentioned SUSY reduction mechanism is generalized
as follows.
{\it{Unless all components in a multiplet of some symmetry group $G$ 
have a common $Z_2$ parity on $S^1/Z_2$,
the symmetry is not preserved after the integration of the fifth
dimension because zero modes, in general, do not form a full 
multiplet of $G$.}}
We apply this mechanism to a reduction of $SU(5)$ gauge symmetry 
in the grand unified theory (GUT).

This paper is organized as follows. 
In the next section, we explain the mechanism of symmetry change 
on $S^1/Z_2$.
For 5D $SU(5)$ GUT with minimal particle content,
we discuss the reduction of gauge symmetry, the mass spectrum 
in 4D theory,
and its phenomenological implications in $\S$3. 
Section 4 is devoted to conclusions and discussion.

\section{Symmetry transition upon compactification}

The space-time is assumed to be factorized into a product of 
4D Minkowski space-time $M^4$ and the orbifold $S^1/Z_2$,
whose coordinates are denoted
by $x^\mu$ ($\mu = 0,1,2,3$) and $y(=x^5)$, respectively.
The 5D notation $x^M$ ($M = 0,1,2,3,5$) is also used.
The orbifold $S^1/Z_2$ is obtained by dividing a circle $S^1$
with radius $R$ with a $Z_2$ transformation which acts on $S^1$
by $y \to -y$.
This compact space is regarded as an interval with a distance of $\pi R$.
There are two 4D walls placed at
fixed points $y=0$ and $y=\pi R$ on $S^1/Z_2$.

An intrinsic $Z_2$ parity of the 5D bulk field $\phi(x^\mu, y)$
is defined by the transformation
\begin{eqnarray}
  \phi(x^\mu, y) \to \phi(x^\mu, -y) = P \phi(x^\mu, y) .
\label{P-tr}
\end{eqnarray}
The Lagrangian should be invariant under the $Z_2$ transformation.
By definition, $P$ equals 1 or $-1$.
We denote the field with $P=1$ ($P=-1$) by $\phi_{+}$ ($\phi_{-}$).
The fields $\phi_{+}$ and $\phi_{-}$ are Fourier expanded as
\begin{eqnarray}
  \phi_{+} (x^\mu, y) &=& 
       {1 \over \sqrt{\pi R}} 
      \sum_{n=0}^{\infty} \phi^{(n)}_{+}(x^\mu) \cos{ny \over R}  ,
\label{phi+exp}\\
  \phi_{-} (x^\mu, y) &=& 
       {1 \over \sqrt{\pi R}}
      \sum_{n=1}^{\infty} \phi^{(n)}_{-}(x^\mu) \sin{ny \over R}  ,
\label{phi-exp}
\end{eqnarray}
where $n$ is an integer, and the fields $\phi^{(n)}_{\pm}(x^\mu)$ 
acquire mass $n/R$ upon compactification. 
Note that 4D massless fields are absent in $\phi_{-}(x^\mu, y)$.

Let us study the case in which a field $\Phi(x^\mu, y)$ is an $N$-plet
under some symmetry group $G$.
Each component of $\Phi$ is denoted by $\phi_k$, i.e.,
$\Phi = (\phi_1, \phi_2, ..., \phi_N)^T$.
The $Z_2$ transformation of $\Phi$ is given by the same form as 
(\ref{P-tr}), but in this case $P$ is an $N \times N$ matrix
\footnote{$P$ is a unitary and hermitian matrix.} 
which satisfies $P^2 = I$, where $I$ is the unit matrix.
The $Z_2$ invariance of the Lagrangian
does not necessarily require that $P$ be $I$ or $-I$.
Unless all components of $\Phi$ have common $Z_2$ parity (i.e.,
if $P \neq \pm I$), a symmetry transition occurs upon compactification
because of the lack of zero modes in components with odd parity.

Here we give two simple examples of a symmetry transition.\\
(a) Scalar field with $N_f$ flavors

We consider a 5D Lagrangian density given by
\begin{eqnarray}
 {\cal L}^{(5)}_S &\equiv&  |\partial_M \Phi|^2 = \sum_{k=1}^{N_f} 
  |\partial_M \phi_k|^2  ,
\label{5DLS}
\end{eqnarray} 
where $\Phi(x^\mu, y)$ is a 5D complex scalar field with $N_f$ components.
The Lagrangian is invariant
under the transformation
$\Phi(x^\mu, y) \to \Phi(x^\mu, -y) = P \Phi(x^\mu, y)$, with $P^2 = I$.
If we take $P={\rm diag}(1,...,1,-1,...,-1)$ where 
the first $M_f$ elements equal 1, 
we derive the following 4D Lagrangian density after integrating out
the fifth dimension and rescaling the 4D fields $\phi_k^{(n)}(x^\mu)$:
\begin{eqnarray}
 {\cal L}^{(4)}_S &=&  \sum_{k=1}^{M_f} |\partial_\mu \phi_k^{(0)}|^2 
 + \sum_{n=1}^{\infty} \sum_{k=1}^{N_f} 
\left(|\partial_\mu \phi_k^{(n)}|^2 
 + {n^2 \over R^2}|\phi_k^{(n)}|^2 \right) .
\label{4DLs}
\end{eqnarray} 
The global symmetry $U(N_f)$ in 5D theory is reduced to
its subgroup $U(M_f)$ upon compactification.
Note that the Kaluza-Klein excitations $\phi_k^{(n)}(x^\mu)$
form a full multiplet of $U(N_f)$.\\
(b) Dirac fermion with $N_f$ flavors

We consider a 5D Lagrangian density given by
\begin{eqnarray}
 {\cal L}^{(5)}_D &=&  i \bar{\Psi} \gamma^M \partial_M \Psi
   = \sum_{k=1}^{N_f} i \bar{\psi}_k \gamma^M \partial_M \psi_k  ,
\label{5DLD}
\end{eqnarray} 
where $\Psi = (\Psi_L, \Psi_R)^T$ is a 5D Dirac fermion 
with $N_f$ components denoted by $\psi_k$.
The Lagrangian is invariant
under the transformation
$\Psi_L(x^\mu, y) \to \Psi_L(x^\mu, -y) = P \Psi_L(x^\mu, y)$ 
and $\Psi_R(x^\mu, y) \to \Psi_R(x^\mu, -y) = -P \Psi_R(x^\mu, y)$,
with $P^2 = I$.
When we take $P={\rm diag}(1,...,1,-1,...,-1)$, where 
the first $M_f$ elements equal 1, 
we obtain the 4D Lagrangian density 
\begin{eqnarray}
 {\cal L}^{(4)}_D &=&  \sum_{k=1}^{M_f} i {\psi}_{Lk}^{(0)\dagger} 
  \sigma^\mu \partial_\mu \psi_{Lk}^{(0)}
  + \sum_{k=1}^{N_f-M_f} i {\psi}_{Rk}^{(0)\dagger}
 \bar{\sigma}^\mu \partial_\mu \psi_{Rk}^{(0)} 
\nonumber \\
&~& + \sum_{n=1}^{\infty} \sum_{k=1}^{N_f} \left(i \bar{\psi}_{k}^{(n)} 
  \gamma^\mu \partial_\mu \psi_{k}^{(n)} 
- {n \over R} \bar{\psi}_{k}^{(n)} \psi_{k}^{(n)} \right)  ,
\label{4DLD}
\end{eqnarray} 
after integrating out the fifth dimension and rescaling 
the 4D fields $\psi_k^{(n)}(x^\mu)$.
Here the components of $\Psi_L$ and $\Psi_R$ are denoted by $\psi_{Lk}$
and $\psi_{Rk}$, respectively.
The 5D theory has the global symmetry $SU(N_f)_V \times U(1)_V$, but
a symmetry in 4D theory
turns out to be $SU(M_f)_L \times SU(N_f-M_f)_R
\times U(1)_V \times U(1)_A$
in the decoupling limit of Kaluza-Klein modes,
which form a full multiplet of $SU(N_f)_V \times U(1)_V$.

\section{A model with SU(5) gauge symmetry}

We apply the symmetry transition mechanism 
to 5D $SU(5)$ GUT with minimal particle content.
We assume that the 5D gauge boson $A_M(x^\mu,y)$ 
and the Higgs boson $\Phi(x^\mu,y)$
exist in the bulk $M^4 \times S^1/Z_2$.
The fields $A_M$ and $\Phi$ form an adjoint representation 
${\bf 24}$ and a fundamental representation ${\bf 5}$ of $SU(5)$,
respectively.
We assume that our visible world is one of 4D walls
(We choose the wall fixed at $y=0$ as the visible one and call it wall I)
and that three families of quarks and leptons,
$3\{\psi_{\bar 5} + \psi_{10}\}$, are located on wall I.
That is, matter fields contain no excited states along the
$S^1/Z_2$ direction.

The gauge invariant action is given by
\begin{eqnarray}
 S &=& \int  {\cal L}^{(5)} d^5x + \int {\cal L}^{(4)} d^4x , \\
 {\cal L}^{(5)} &\equiv&  - {1 \over 2} {\rm tr} F_{MN}^2 + 
|D_M \Phi|^2 - V(\Phi) ,\\
{\cal L}^{(4)} &\equiv&  \sum_{3 {\rm families}} ( i \bar{\psi}_{10} 
\gamma^\mu D_\mu {\psi}_{10}
    + i \bar{\psi}_{\bar 5} \gamma^\mu D_\mu {\psi}_{\bar 5} 
\nonumber\\
 &~& + f_{U(5)} \Phi \psi_{10} \psi_{10} 
 + f_{D(5)} \Phi^{\dagger} \psi_{10} \psi_{\bar 5} 
 + \mbox{h.c.} )  ,
\end{eqnarray}
where $D_M \equiv \partial_M - i g_{(5)} A_M(x^\mu, y)$,
$g_{(5)}$ is a 5D gauge coupling constant, and $f_{U(5)}$ and $f_{D(5)}$ 
are 5D Yukawa coupling matrices.
The representations of $\psi_{\bar{5}}$ and $\psi_{10}$ 
are $\bar{\bf 5}$ and ${\bf 10}$ under $SU(5)$, respectively.
In ${\cal L}^{(4)}$, the bulk fields $A_\mu$ and $\Phi$ are replaced
by fields with values at wall I, $A_\mu(x^\mu, 0)$ 
and $\Phi(x^\mu, 0)$.
The Lagrangian is invariant under
the $Z_2$ transformation
\begin{eqnarray}
 &~& A_{\mu}(x^\mu, y) \to A_{\mu}(x^\mu, -y) = 
P A_{\mu}(x^\mu, y) P^{-1} , \nonumber \\
 &~& A_{5}(x^\mu, y) \to A_{5}(x^\mu, -y) = 
- P A_{5}(x^\mu, y) P^{-1} , \nonumber \\
 &~& \Phi(x^\mu, y) \to \Phi(x^\mu, -y) = P \Phi(x^\mu, y)  .
\label{P-tr2}
\end{eqnarray}

When we take $P={\rm diag}(-1,-1,-1,1,1)$,
the $SU(5)$ gauge symmetry is reduced to that of
the Standard Model, $G_{SM} \equiv SU(3) \times SU(2)
\times U(1)$, in 4D theory.\footnote{
Our symmetry reduction mechanism is different from the Hosotani mechanism.
In fact, the Hosotani mechanism does not work in our case,
because $A_{5}^{a}(x^\mu,y)$ has odd parity, as given in (\ref{P-tr2}),
and its VEV should vanish.}
This is because the boundary conditions on $S^1/Z_2$ 
given in (\ref{P-tr2}) do not respect $SU(5)$ symmetry,
as we see from the relations for the gauge generators $T^{\alpha}$ 
$(\alpha = 1, 2,...,24)$,
\begin{eqnarray}
  P T^a P^{-1} = T^a , ~~
  P T^{\hat{a}} P^{-1} = -T^{\hat{a}}  .
\end{eqnarray}
The $T^a$s are gauge generators of $G_{SM}$
and the $T^{\hat{a}}$s are other gauge generators.\footnote{
We expect that the specific $Z_2$ parity given by 
$P={\rm diag}(-1,-1,-1,1,1)$
is determined non-perturbatively in an underlying theory.}
After integrating out the fifth dimension,
we obtain the 4D lagrangian density
\begin{eqnarray}
{\cal L}^{(4)}_{\rm eff} &=& {\cal L}^{(4)}_B + {\cal L}^{(4)} ,\\
{\cal L}^{(4)}_B &\equiv& - {1 \over 4} \sum_a {F_{\mu\nu}^{a(0)}}^2
  + |D_\mu A_5^{\hat{a}(0)}|^2 \nonumber \\
&~& + |D_\mu \phi_W^{(0)}|^2 + g_U^2|A_5^{\hat{a}(0)} \phi_W^{(0)}|^2
 - V(\phi_W^{(0)}) + \cdots , \\
{\cal L}^{(4)} &\equiv& \sum_{3 {\rm families}} 
[ i \bar{\psi}_{10} \gamma^\mu 
(\partial_\mu - i g_{U} \sum_{n=0}^{\infty} 
  A_{\mu}^{a(n)} T^a({\bf 10})) {\psi}_{10} \nonumber \\
&~&    + i \bar{\psi}_{\bar 5} \gamma^\mu
(\partial_\mu - i g_{U} \sum_{n=0}^{\infty} 
  A_{\mu}^{a(n)} T^a(\bar{\bf 5})) {\psi}_{\bar 5} 
\nonumber\\
 &~& + f_U \sum_{n=0}^{\infty} \phi_W^{(n)} q {\bar u} 
 + f_D \sum_{n=0}^{\infty} \tilde{\phi}_W^{(n)} q {\bar d} 
 + f_D \sum_{n=0}^{\infty} \tilde{\phi}_W^{(n)} l {\bar e} 
 + \mbox{h.c.} 
\nonumber\\
 &~& + \sum_{n=0}^{\infty} ( g_U \bar{\psi}_{10} \gamma^5 
A_{5}^{\hat{a}(n)} T^{\hat{a}}({\bf 10}) {\psi}_{10} 
+ g_U \bar{\psi}_{\bar 5} \gamma^5
A_{5}^{\hat{a}(n)} T^{\hat{a}}(\bar{\bf 5}) {\psi}_{\bar 5}) ]  ,
\label{4D-L}
\end{eqnarray}
where the dots in ${\cal L}^{(4)}_B$ represent terms including 
Kaluza-Klein modes, $g_U$ $(\equiv g_{(5)}/\sqrt{\pi R})$ is a 4D 
gauge coupling constant, $f_U$ $(\equiv f_{U(5)}/\sqrt{\pi R})$ 
and $f_D$ $(\equiv f_{D(5)}/\sqrt{\pi R})$ are 4D Yukawa
coupling matrices, $q$, $\bar{u}$ and $\bar{d}$ are quarks, 
$l$ and $\bar{e}$ are leptons, and $\phi_W$ 
($\tilde{\phi}_W^{(n)} \equiv i \tau_2 \phi_W^{(n)*}$)
is a weak Higgs doublet.
The massive modes should be rescaled by a factor of $\sqrt{2}$, e.g.,
$A_{\mu}^{a(n)} \to \sqrt{2} A_{\mu}^{a(n)}$ and 
$\phi_W^{(n)} \to \sqrt{2} \phi_W^{(n)}$ $(n \neq 0)$,
as a result of the proper normalization of their kinetic terms.
The mass spectrum after compactification is given in Table I.
\begin{table}[b]
\caption{Mass spectrum at the tree level.}
\begin{center}
\begin{tabular}{l|l|l}
\hline\hline
4D fields & Quantum numbers & Mass \\
\hline
$A_{\mu}^{a(0)}(x^\mu)$ & $({\bf 8}, {\bf 1}) + ({\bf 1}, {\bf 3})
 + ({\bf 1}, {\bf 1})$ & 0 \\
$A_{5}^{\hat{a}(0)}(x^\mu)$ & $({\bf 3}, {\bf 2}) + (\bar{\bf 3}, {\bf 2})$
 & 0 \\
$\phi_W^{(0)}(x^\mu)$ & $({\bf 1}, {\bf 2})$ & $\sqrt{\mu^2}$ \\
\hline
$A_{M}^{\alpha(n)}(x^\mu)$ & $({\bf 8}, {\bf 1}) + ({\bf 1}, {\bf 3})
 + ({\bf 1}, {\bf 1})$ & $n/R$ \\
$(n \neq 0)$ & $~~+ ({\bf 3}, {\bf 2}) + (\bar{\bf 3}, {\bf 2})$ & ~ \\
$\Phi^{(n)}(x^\mu)$ & $({\bf 3}, {\bf 1}) + ({\bf 1}, {\bf 2})$ 
& $\sqrt{(n/R)^2 + \mu^2}$ \\
$(n \neq 0)$ &  & \\
\hline
$\psi_{\bar{5}}(x^\mu)$ & $(\bar{\bf 3}, {\bf 1}) + ({\bf 1}, {\bf 2})$
 & 0 \\
$\psi_{10}(x^\mu)$ & $({\bf 3}, {\bf 2}) + (\bar{\bf 3}, {\bf 1})
+ ({\bf 1}, {\bf 1})$ & 0 \\
\hline
\end{tabular}
\end{center}
\end{table}
In the second column, we give $SU(3) \times SU(2)$ quantum numbers
of 4D fields.
The triplet-doublet mass splitting of the Higgs boson is realized
by projecting out zero modes of colored components in the Higgs boson.
There exist extra 4D scalar fields 
$A_{5}^{\hat{a}(0)}(x^\mu)$,
whose quantum numbers are $({\bf 3}, {\bf 2}, -5/6) + 
(\bar{\bf 3}, {\bf 2}, 5/6)$ under $G_{SM}$, and they couple to
$G_{SM}$ gauge bosons, a weak Higgs doublet, and matter fermions.
There is a possibility that $A_{5}^{\hat{a}(0)}$ becomes
superheavy and that the proton decay induced by the exchange of
$A_{5}^{\hat{a}(0)}$ is suppressed, 
while $\phi^{(0)}_{W}$ remains in the weak scale spectrum
after radiative corrections are received.
This is based on the following premises.
The field $A_{5}^{\hat{a}(0)}$ has a vanishing bare mass because of
gauge symmetry, but $\phi^{(0)}_{W}$ has a mass whose square is given by 
\begin{eqnarray}
\mu^2 = {\partial^2 V \over \partial \phi^{(0)2}_W} .
\end{eqnarray}
Both $A_{5}^{\hat{a}(0)}$ and $\phi^{(0)}_{W}$ receive radiative
corrections.
The renormalized mass of $\phi^{(0)}_{W}$ is on the order of the weak
scale.

The theory predicts that coupling constants are unified 
around the compactification scale $M_C (\equiv 1/R)$,
as in the ordinary $SU(5)$ GUT,\cite{GUT}
\begin{eqnarray}
&~& g_3 = g_2 = g_1 = g_U , ~~ f_d = f_e = f_D~,
\end{eqnarray}
where $f_d$ and $f_e$ are Yukawa coupling matrices on
down-type quarks and electron-type leptons, respectively.\footnote{
The unification conditions are, in general, corrected
by non-renormalizable interactions.}
The other feature is that
quarks and leptons couple neither to $X$ and $Y$ gauge bosons 
$A_{\mu}^{\hat{a}(n)}$
nor to the colored Higgs triplet $\phi_C^{(n)}$ at the tree level.
Hence it is expected that the proton decay process 
due to $X$ and $Y$ gauge bosons $A_{\mu}^{\hat{a}(n)}$
is suppressed.\footnote{
If we consider non-renormalizable interactions, 
there are dangeorous
interactions including derivatives with respect to the fifth coordinate, 
which can induce a rapid proton decay.}

\section{Conclusions and discussion}

We have studied a mechanism of symmetry transition
upon compactification of 5D field theory on $S^1/Z_2$.
The transition occurs unless all components in a multiplet 
of a symmetry group have a common $Z_2$ parity
on $S^1/Z_2$.
This mechanism has been applied to a reduction of gauge symmetry
in 5D $SU(5)$ GUT.
Under the assumption that our visible world is a 4D wall fixed at
$y=0$ and that quarks and leptons live on the wall, 
we have derived the same type of action as that in the Standard Model.
The triplet-doublet mass splitting on Higgs bosons is realized
at the tree level by the $Z_2$ projection.
In the sector with renormalizable interactions,
the theory predicts the coupling unification
$g_3 = g_2 = g_1 = g_U$ and $f_d = f_e = f_D$.
Quarks and leptons couple neither to off-diagonal gauge bosons 
$A_{\mu}^{\hat{a}}$ nor to the colored Higgs triplet $\phi_C$ 
at the tree level.

On the other hand, there are several problems in our $SU(5)$ model.
Here we list some of them.
The first one is the existence of 4D scalar fields 
$A_{5}^{\hat{a}(0)}(x^\mu)$ with quantum numbers 
$({\bf 3}, {\bf 2}, -5/6) + (\bar{\bf 3}, {\bf 2}, 5/6)$ under $G_{SM}$
in the weak scale spectrum at the tree level.
Their presence induces a dangerous proton decay because they couple
to quarks and leptons at the tree level.
To avoid this problem, 
these scalar fields should be superheavy.
They can acquire superheavy masses by radiative corrections.
Unless the radiative corrections to $A_{5}^{\hat{a}(0)}(x^\mu)$
are finite, the renormalizability can be violated 
because of the lack of mass and self-interaction terms
of $A_{5}^{\hat{a}(0)}(x^\mu)$.
This problem is under investigation based on the 5-dimensional
calculation discussed in Ref.\citen{HIL}.
Otherwise it is necessary to introduce extra Higgs bosons 
in order to eliminate unwanted particles from the low-energy spectrum.
The second problem involves the question of
how to break the electro-weak symmetry naturally
and how to stabilize the weak scale.
Our model suffers from gauge hierarchy problem.\cite{ghp}
The third problem regards the reality of coupling unification, i.e.,
whether or not our model is consistent with experimental data of
gauge couplings and fermion masses.
Power-law corrections from the extra space $S^1/Z_2$ 
should also be considered.\cite{DDG}
The fourth problem concerns the necessity of 
non-universal $Z_2$ parity, i.e.,
whether or not there is a selection rule which picks out
a specific $Z_2$ parity to break $SU(5)$ down to $G_{SM}$.
The last problem regards how matter fields are localized on the 4D wall.
In spite of these problems, it would be worthwhile to
search for a realistic model of grand unification
in this direction.\footnote{
Attempts to construct GUT have been made through the
dimensional reduction over coset space.\cite{coset}}

\section*{Acknowledgements}
This work is supported by a Grant-in-Aid for Scientific Research 
($\sharp$10740111) from the Ministry of
Education, Science and Culture.

\end{document}